\renewcommand{\vec}[1]{\mathbf{#1}}
\documentclass{iau} 
\usepackage{graphicx}

\title[Various scenarios for the equatorward migration of sunspots] %% give here short title %%
{Various scenarios for the equatorward migration of sunspots}
\author[Detlef Elstner, Yori Fournier \& Rainer Arlt]   %% give here short author list %%
{Detlef Elstner$^1,$
  Yori Fournier$^1$ \and Rainer Arlt$^1$}

\affiliation{$^1$Leibniz-Institute for Astrophysics Potsdam, \\ An der Sternwarte 16, 
D-14482 Potsdam, Germany \\ email: {\tt delstner@aip.de} \\[\affilskip]
}

\pubyear{2019}
\volume{354}  %% insert here IAU Symposium No.
\jname{Solar and Stellar Magnetic Fields: Origins and Manifestations}
\editors{A. G. Kosovichev, K. G. Strassmeier \& M. M. Jardine, eds.}
\begin{document}

\maketitle

\begin{abstract}
The profile of the differential rotation together with the sign of the alpha-effect determine the dynamo wave direction. In early models of the solar dynamo the dynamo wave often leads to a poleward migration of the activity belts. Flux transport by the meridional flow or the effect of the surface shear layer are possible solutions. 
In a model including the corona, we show that various migrations can be obtained by varying the properties of the corona.
A new dynamo of Babcock-Leighton type also leads to the correct equatorward migration by the non-linear relation between flux density and rise time of the flux. 
\keywords{solar dynamo, solar corona}
\end{abstract}

\firstsection % if your document starts with a section,
\section{Introduction}

The butterfly diagram of the solar magnetic field is still challenging dynamo theory. 
The simple explanation by a dynamo wave fails because of the increasing angular velocity at lower latitudes.
Including the effect of a meridional circulation leads to the flux transport dynamos, which can explain many features of 
the solar cycle under the assumption of low turbulent diffusivities 
(eg., \cite[Choudhuri et al. 1995]{Choudhuri95}, \cite[Dikpati \& Charbonneau 1999]{Dikpati99}, \cite[Bonnano et al.  2002]{Bonanno02}). 
Most of these models use the mathematically convenient but physically questionable vacuum boundary condition at the solar surface. A better choice is probably 
a force free magnetic field extension into the atmosphere (\cite[Bonanno 2016]{Bonanno16}). 
The role of the corona for the dynamo is still poorly understood. Dynamical models including parts of a corona with high viscosity and diffusivity were investigated by \cite{Warnecke13}. We consider kinematic models with various assumptions for the mean rotation and the turbulent diffusivity in the next section. 
A second possibility for the occurrence of a solar butterfly diagram by nonlocal and finite time correlation effects 
of the turbulence is given in the last section. 

\section{The role of the Corona}

We present  simple 3-dimensional $\alpha^2\Omega$-dynamos  with a solar rotation law neglecting the surface shear layer and
meridional flow. 
We solve the induction equation 
\begin{equation}
{\partial \vec{B} \over \partial t} = 
{\rm curl}( \vec{u} \times \vec{B} + \alpha \circ \vec{B} 
-{\eta}_{\rm T} {\rm {curl}}{\vec{B}}), 
\label{eq:1}
\end{equation} 
in spherical coordinates (${\rm r,\theta,\varphi}$).
The mean flow $\vec{u}=(0,0,\rm r \hspace{0.25em} sin(\theta) \hspace{0.25em} \Omega)$ is  
a solar type rotation within the convection zone similar to \cite{Dikpati99}.

For the rotation of the atmosphere we consider 3 different cases: 
\begin{itemize}
\item[(1)] no radial shear at the stellar surface, same latitudinal dependence as in the convection zone 
\item[(2)] rigid rotation same as the core
\item[(3)] rigid rotation same as the pole at stellar surface
\end{itemize} 

The $\alpha$-tensor has only diagonal components  $\alpha_{ii}=\alpha_0 {\rm cos}(\theta)$ in the convection zone 
independent of solar radius with $\alpha_0=5\hspace{0.25em}{\rm cm/s}$ and is locally quenched with magnetic energy density. 
The diffusivity $\eta_c$ in the convection zone is $5\times10^{11}{\rm cm^2/s}$. 
The inner radial boundary at 0.7 solar radius is a perfect conductor. We add a solar atmosphere up to 1.5 $R_\odot$. 
There we set a pseudo vacuum boundary condition (radial magnetic field only) and consider diffusivities in the atmosphere 
$\eta_h=10\eta_c$, $\eta_h=\eta_c$ and $\eta_h=0.1\eta_c$. 
  
\begin{figure}[!htb]
% \vspace*{-2.0 cm}
\begin{center}
%\hspace{-0.5 cm}
 \includegraphics[width=2.3in]{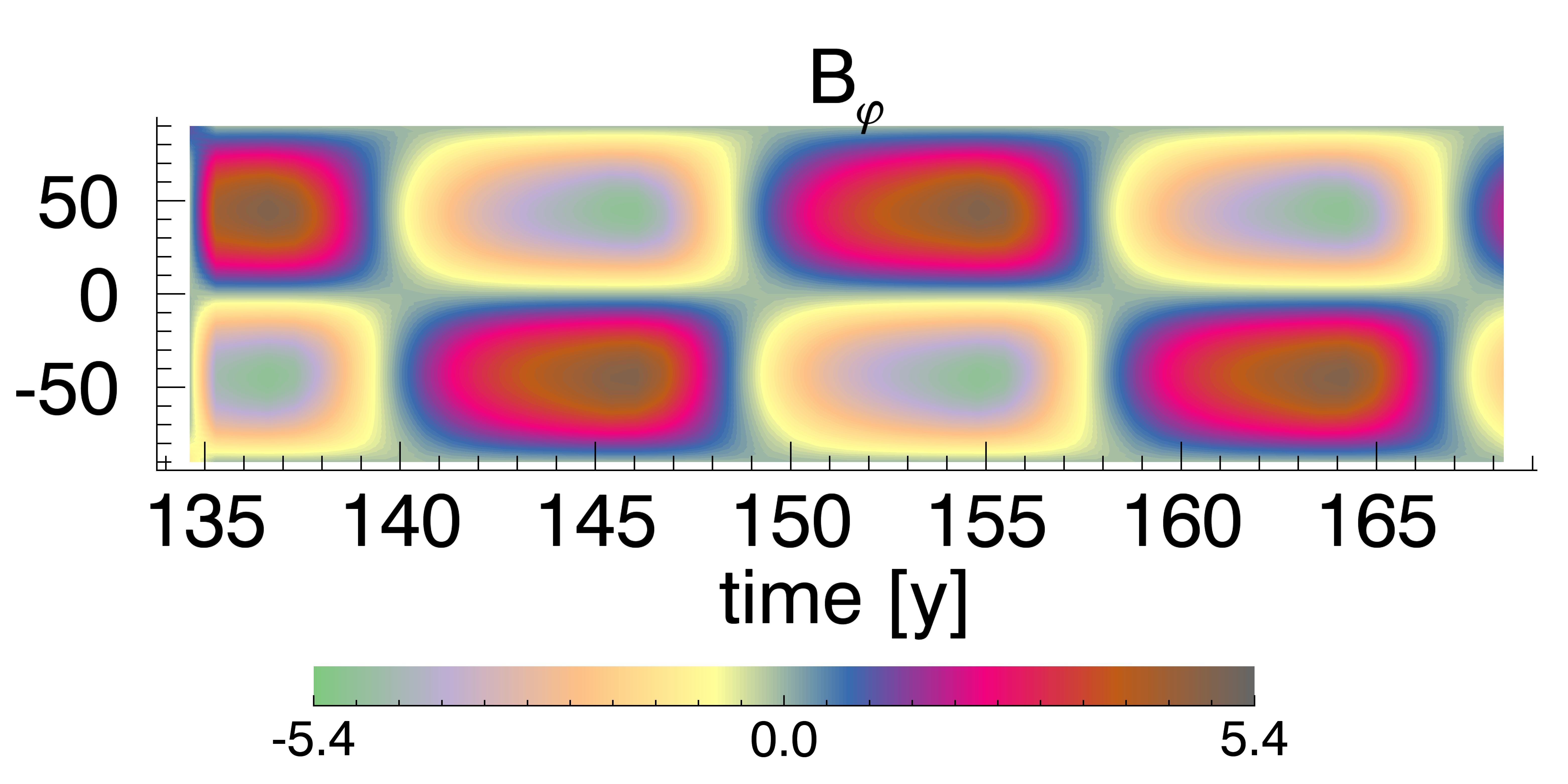} 
 \includegraphics[width=2.3in]{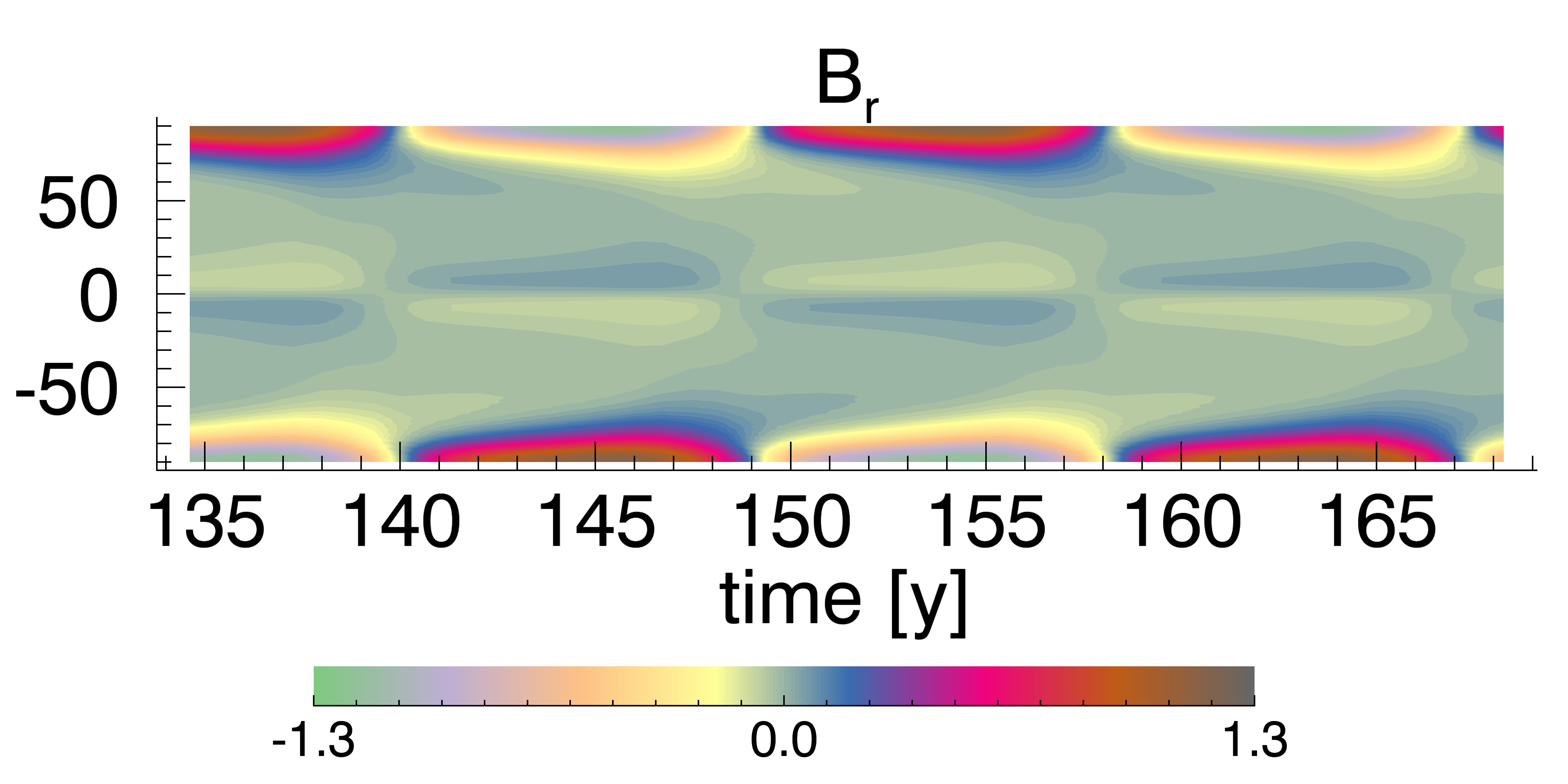} 
 \includegraphics[width=2.3in]{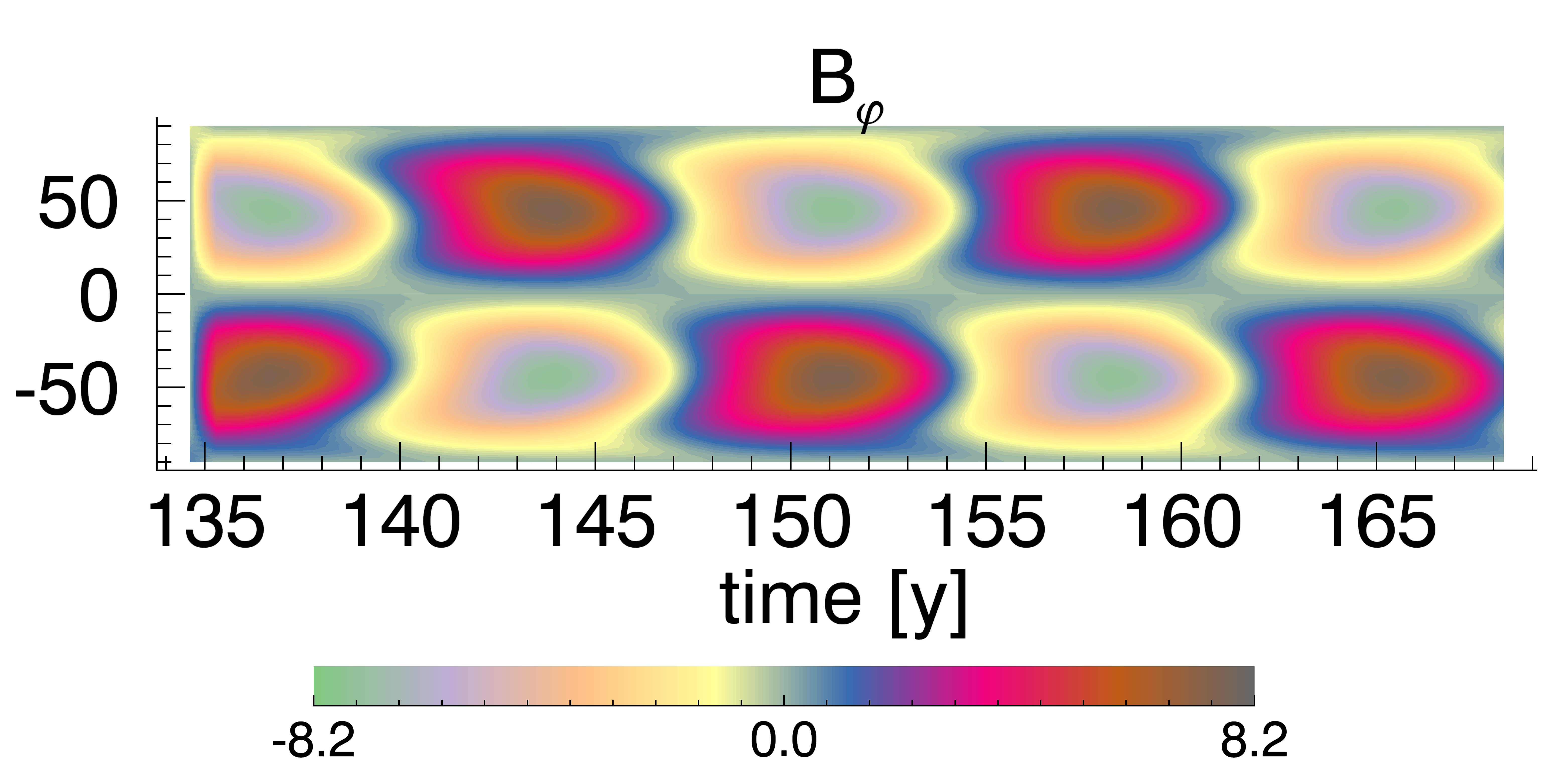}
 \includegraphics[width=2.3in]{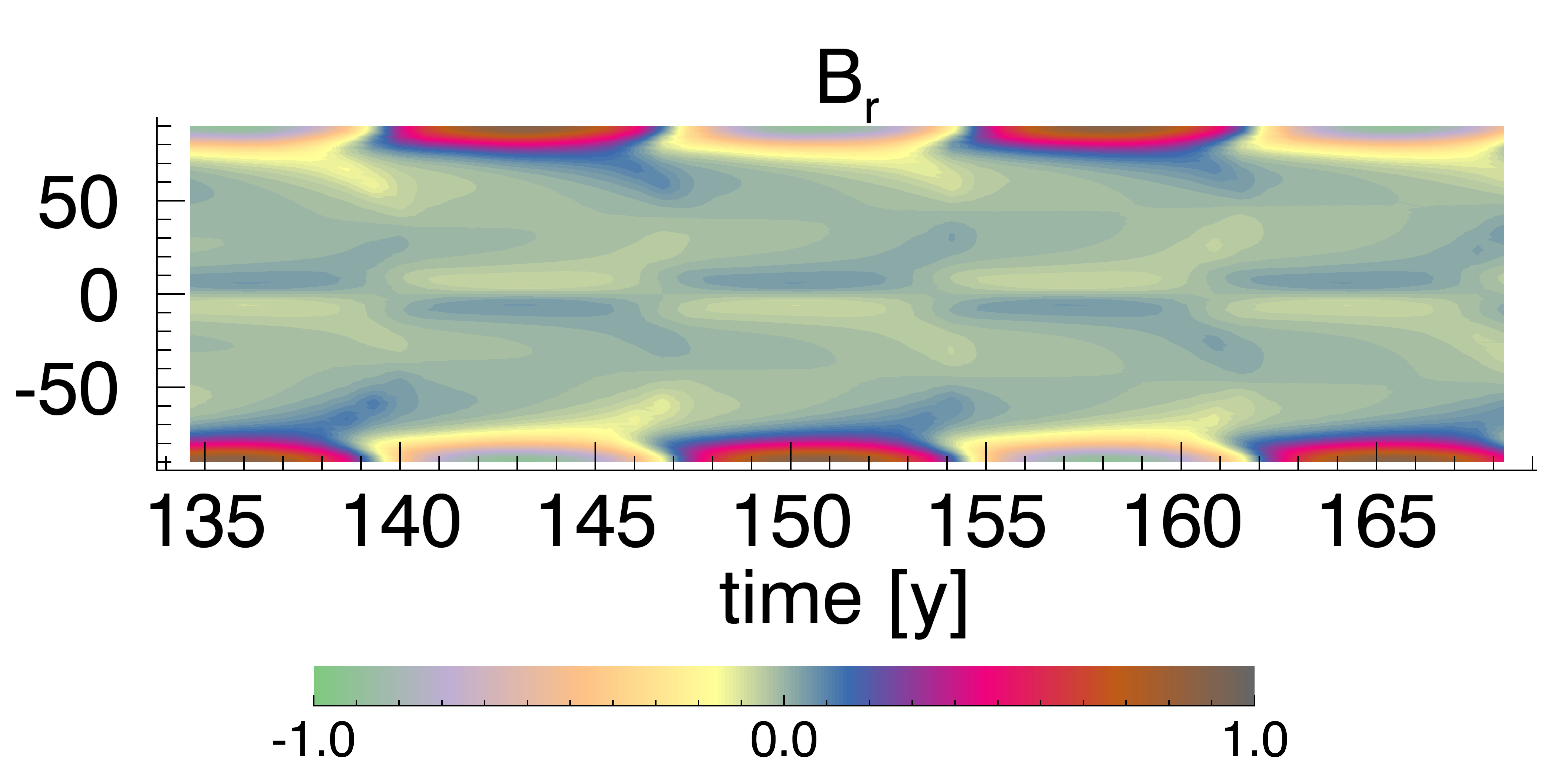}
 \includegraphics[width=2.3in]{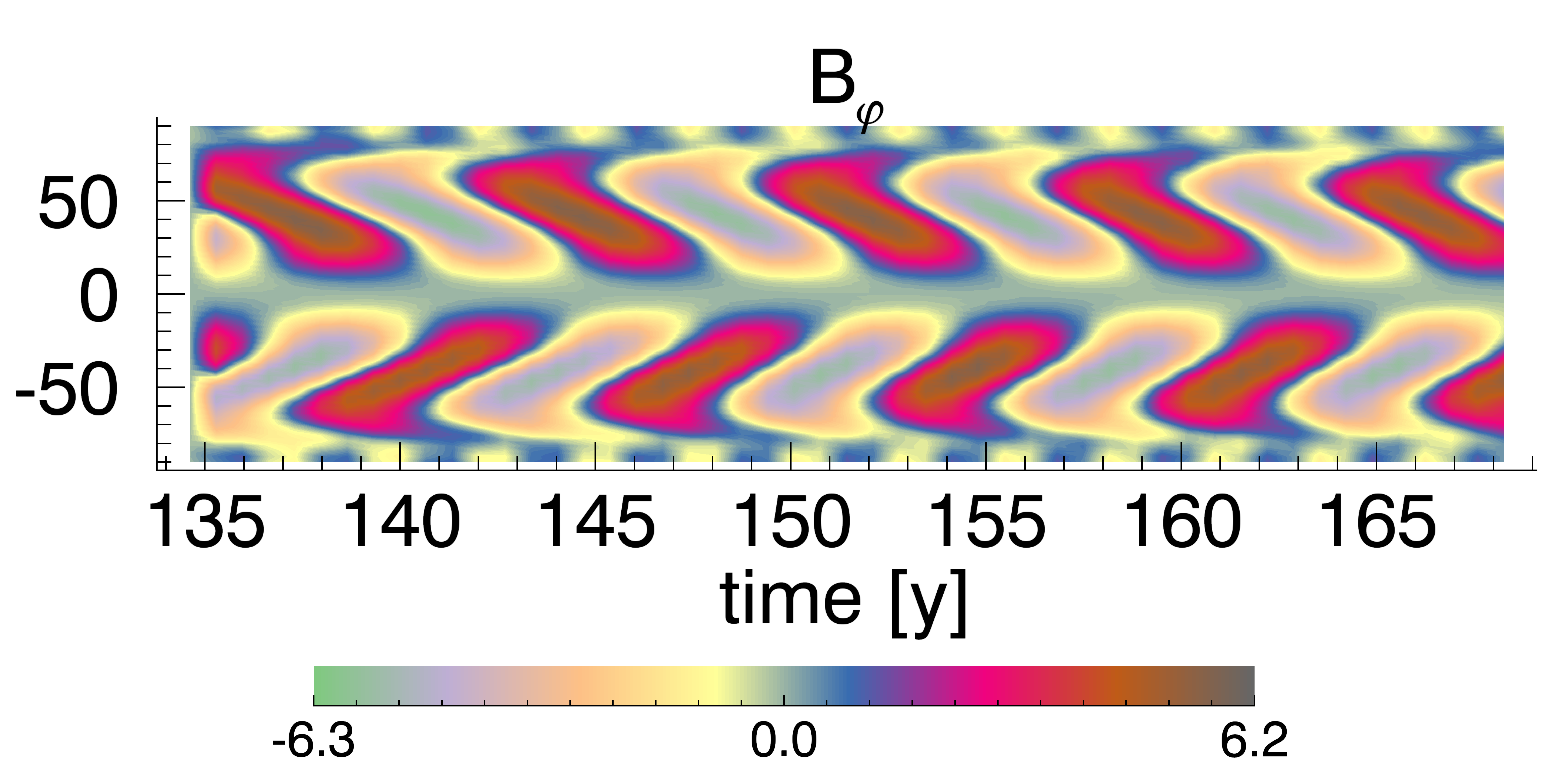}
 \includegraphics[width=2.3in]{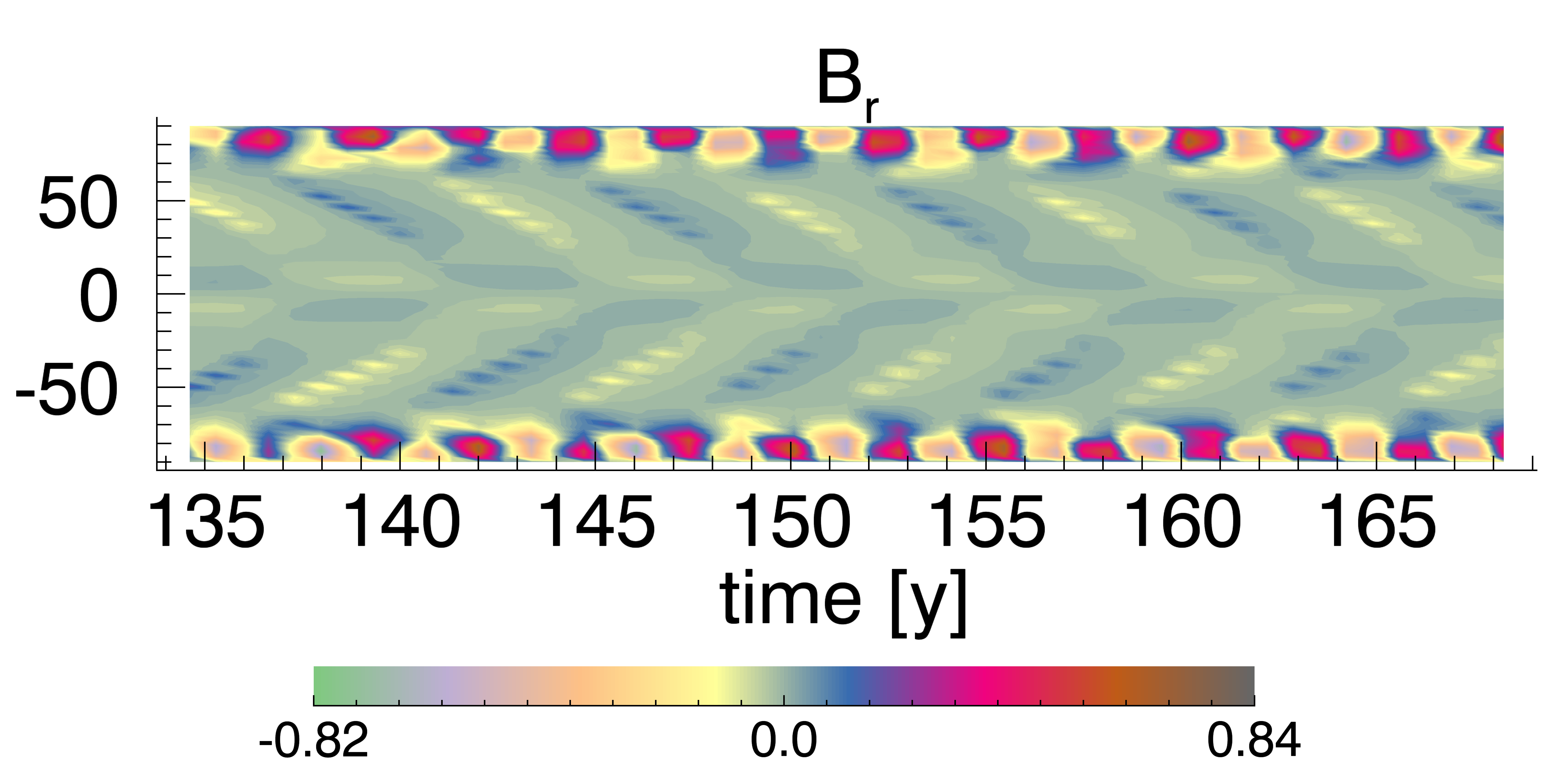}
% \vspace*{-3.0 cm}
 \caption{Time-latitude diagrams for halo diffusivities $\eta_h=10\eta_c$ (upper), $\eta_h=\eta_c$ (middle) and $\eta_h=0.1\eta_c$ (lower). The halo rotates with the same latitudinal dependence as the convection zone.}
   \label{fig1}
\end{center}
\end{figure}
\begin{figure}[!htb]
% \vspace*{-2.0 cm}
\begin{center}
%\hspace{-0.5 cm}
 \includegraphics[width=2.3in]{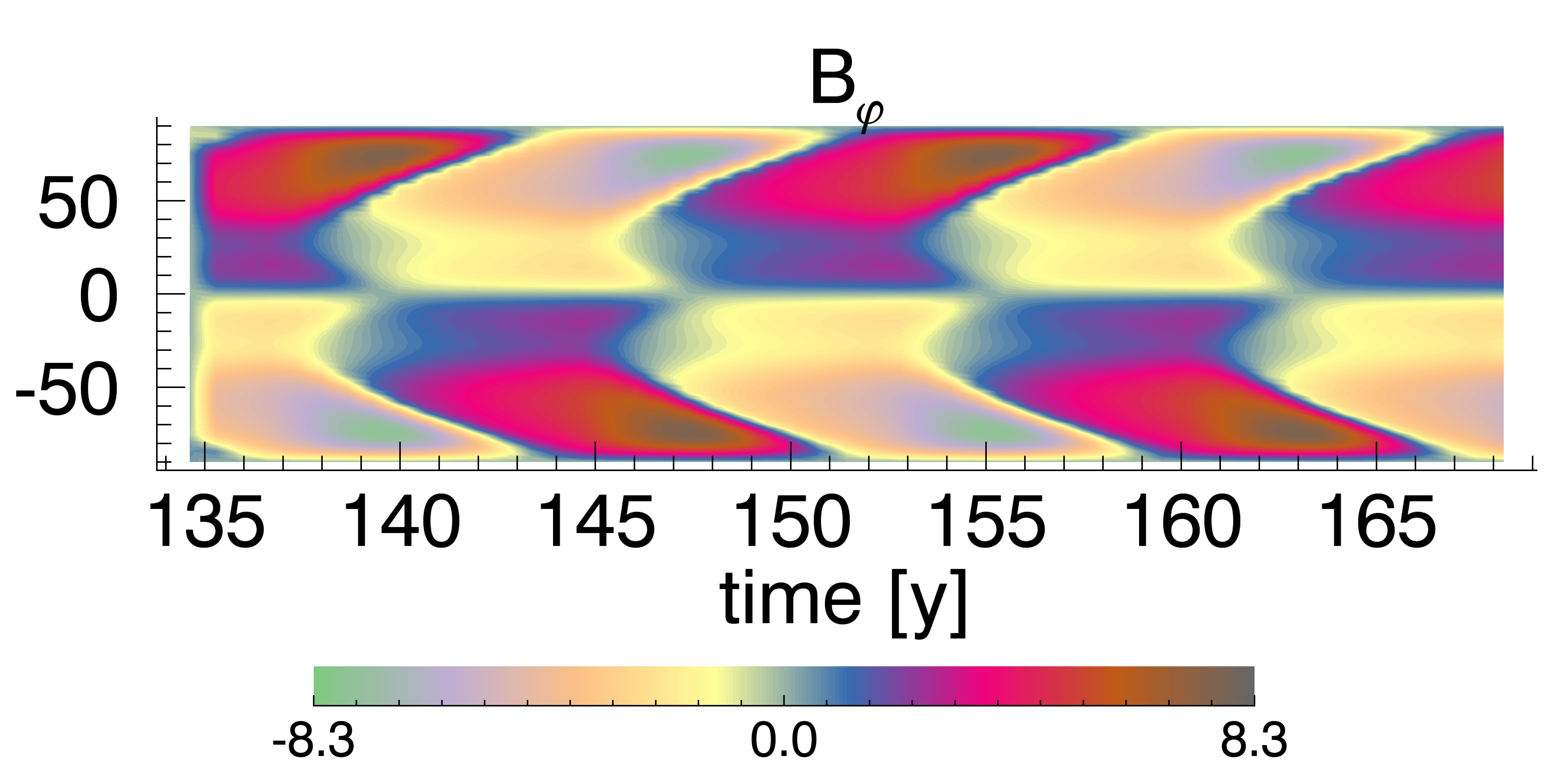} 
 \includegraphics[width=2.3in]{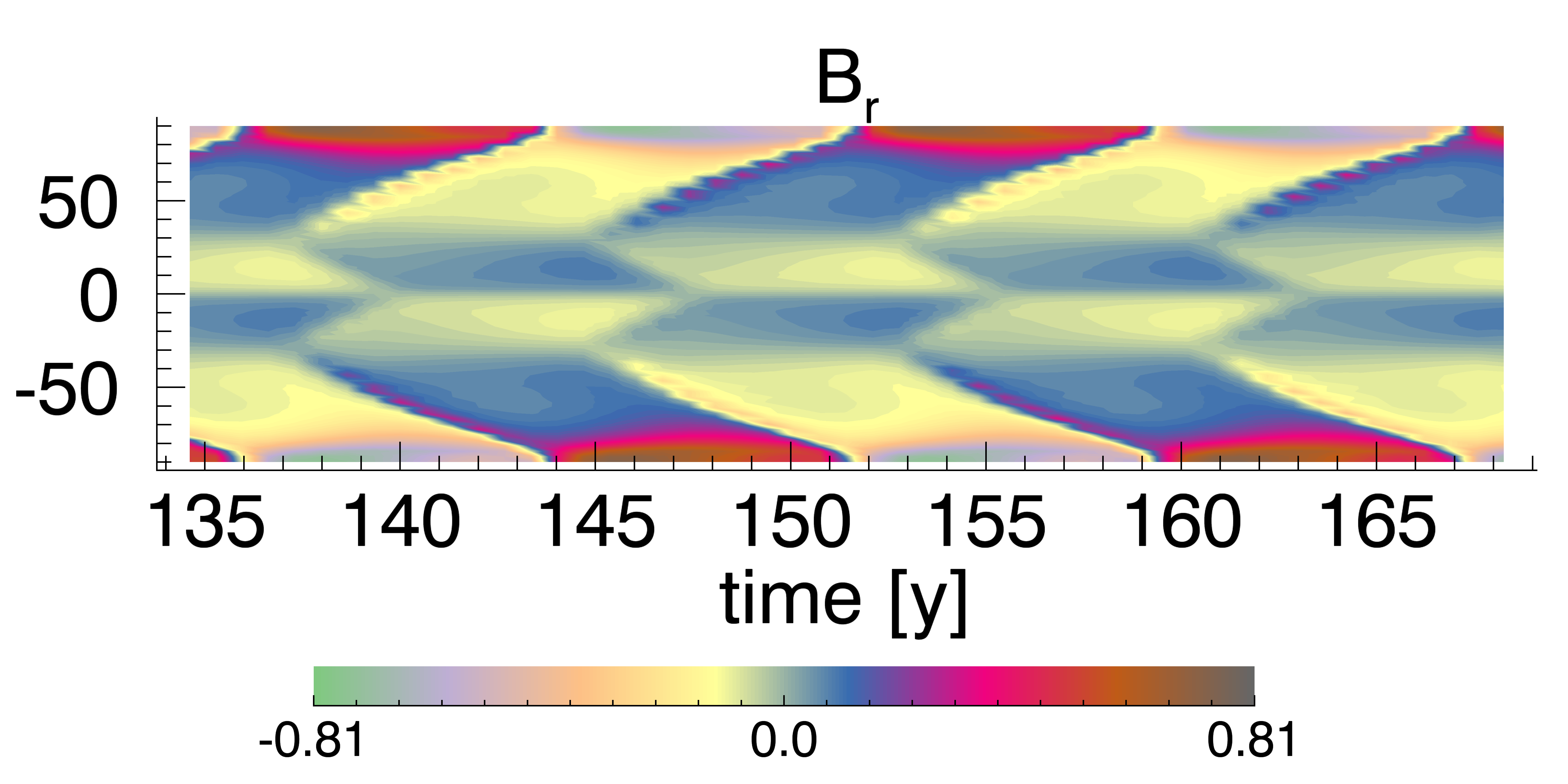} 
 \includegraphics[width=2.3in]{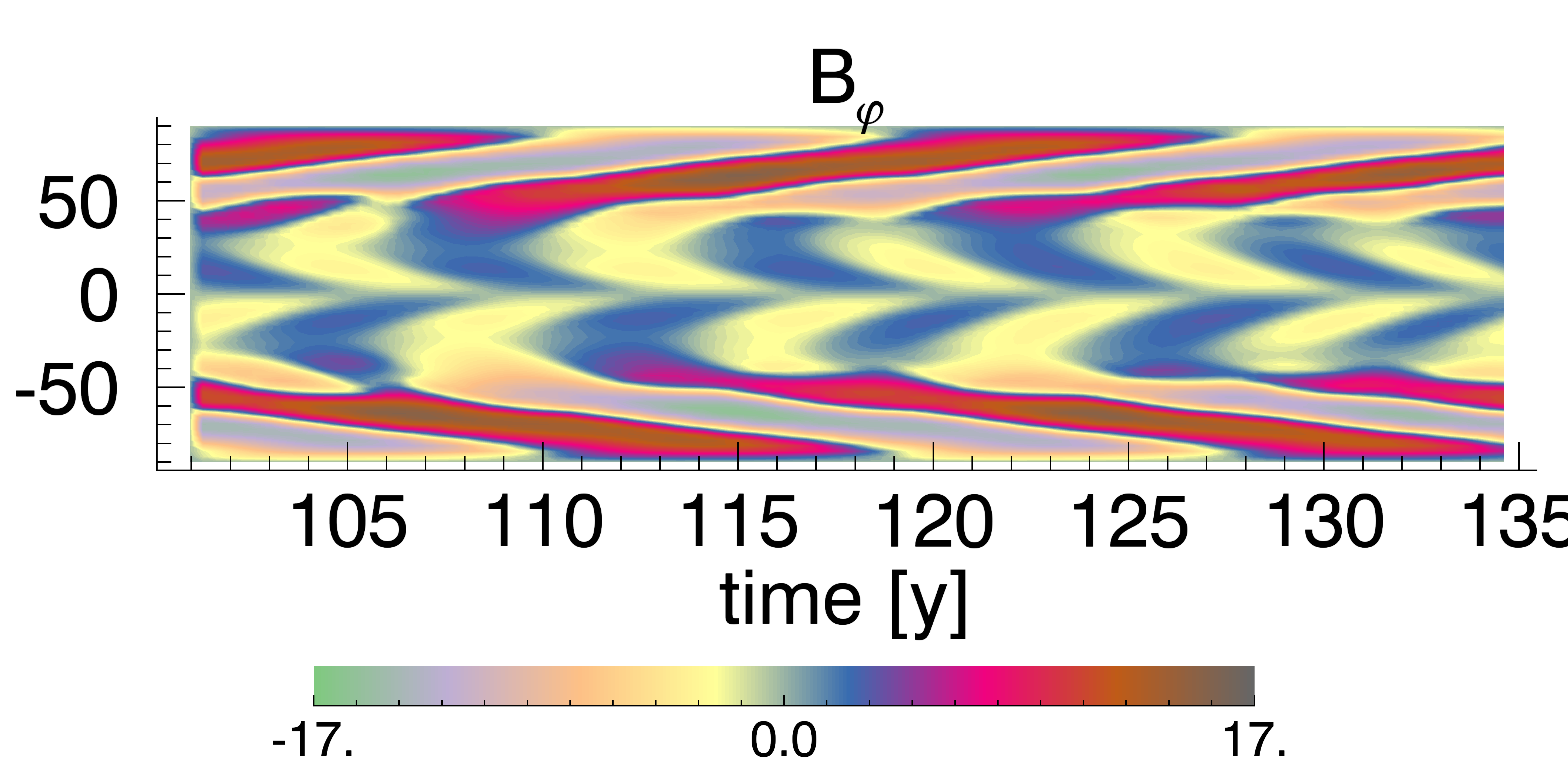}
 \includegraphics[width=2.3in]{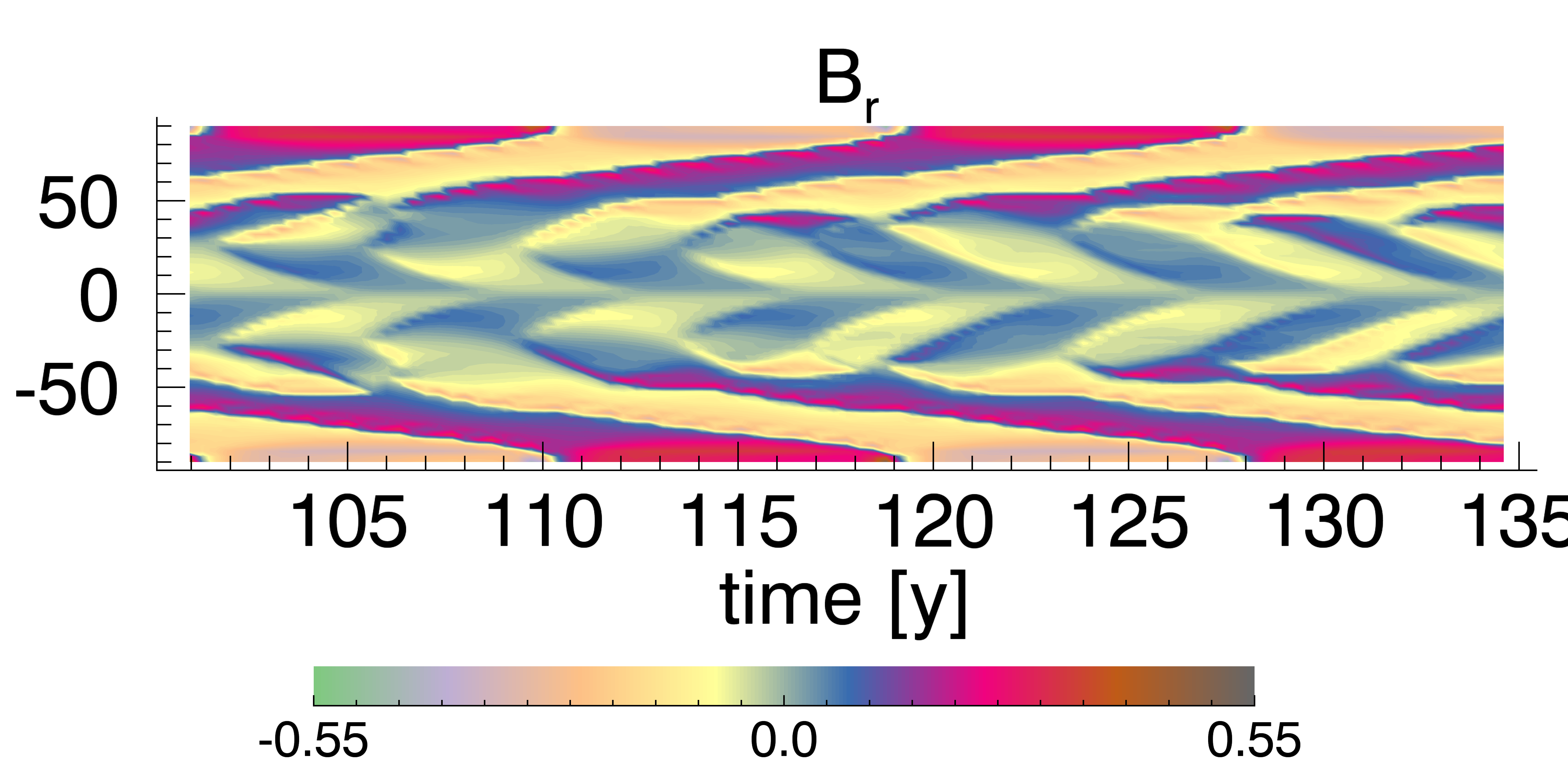}
 \includegraphics[width=2.3in]{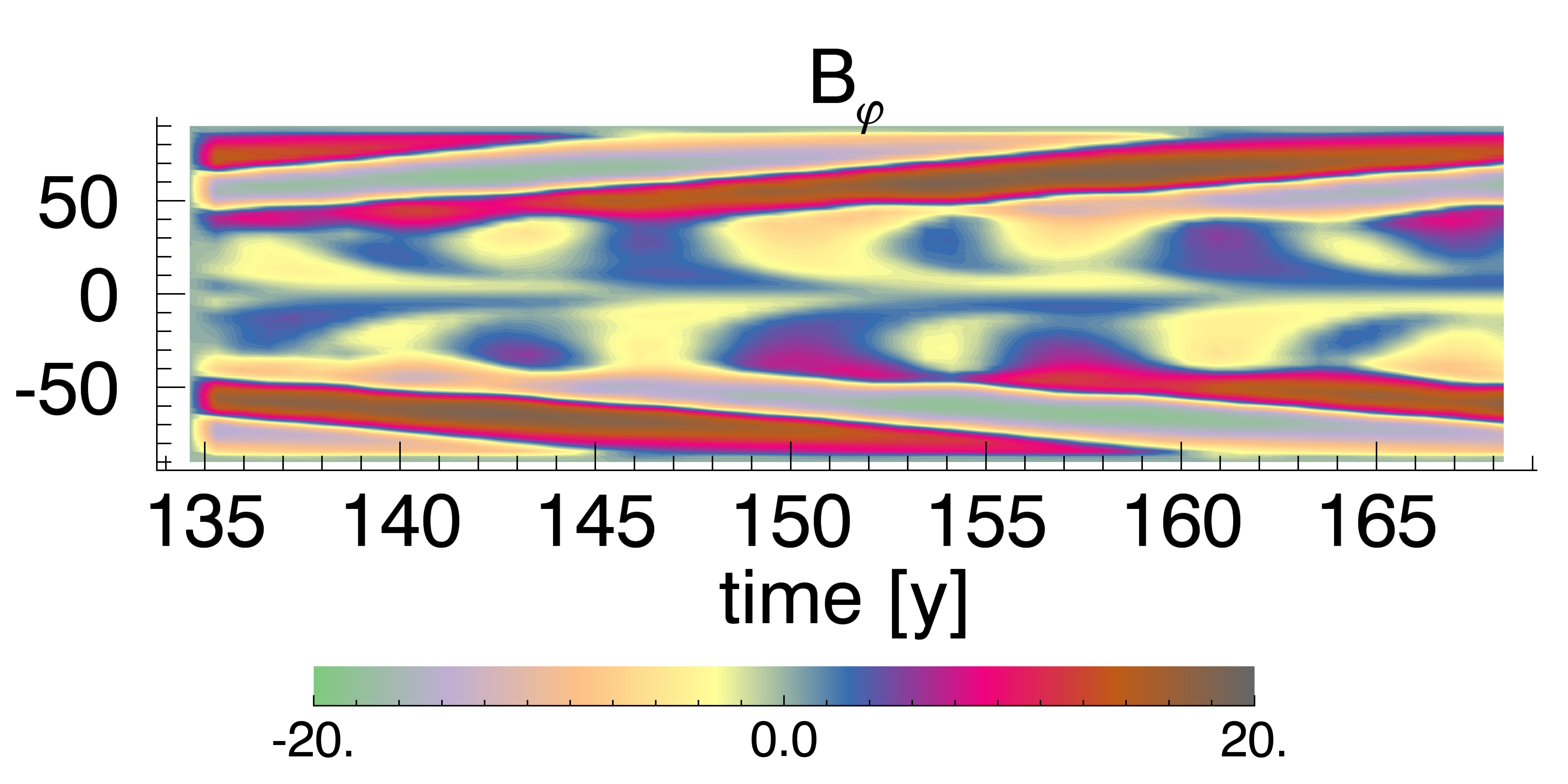}
 \includegraphics[width=2.3in]{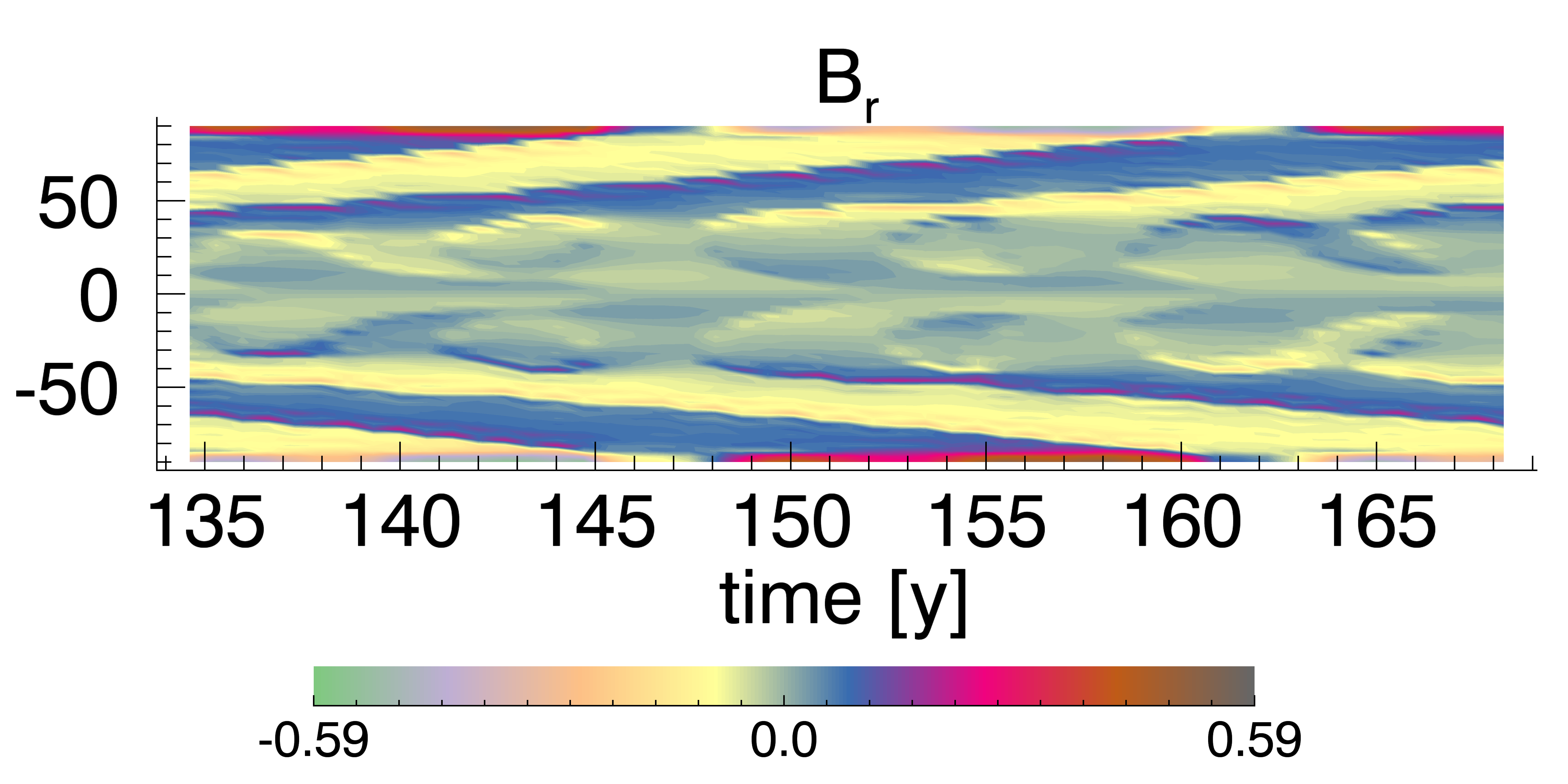}
% \vspace*{-3.0 cm}
 \caption{Time-latitude diagrams for halo diffusivities $\eta_h=10\eta_c$ (upper), $\eta_h=\eta_c$ (middle) and $\eta_h=0.1\eta_c$ (lower). The halo rotates with the core.}
   \label{fig2}
\end{center}
\end{figure}
\begin{figure}[!htb]
% \vspace*{-2.0 cm}
\begin{center}
%\hspace{-0.5 cm}
 \includegraphics[width=2.3in]{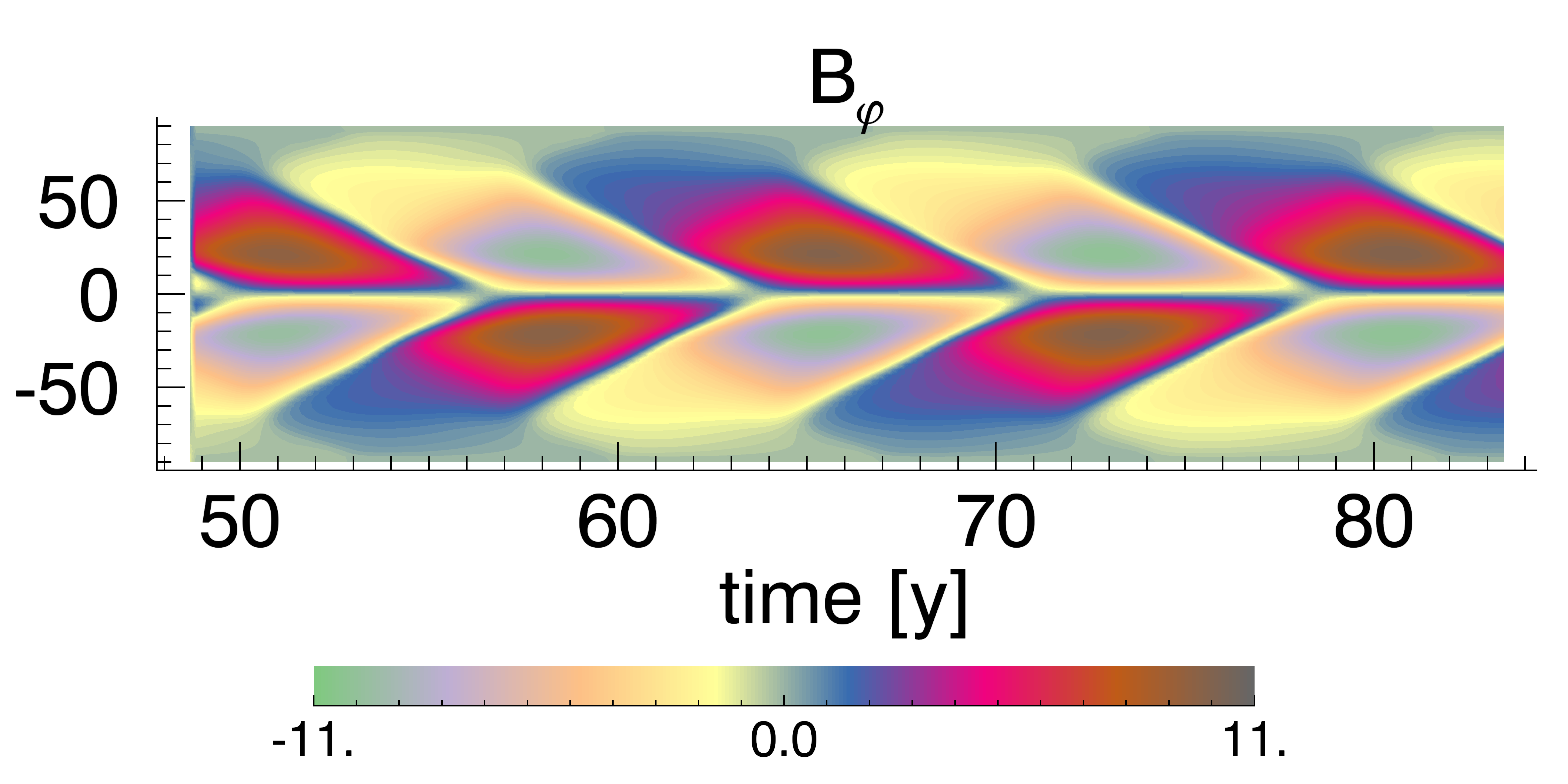} 
  \includegraphics[width=2.3in]{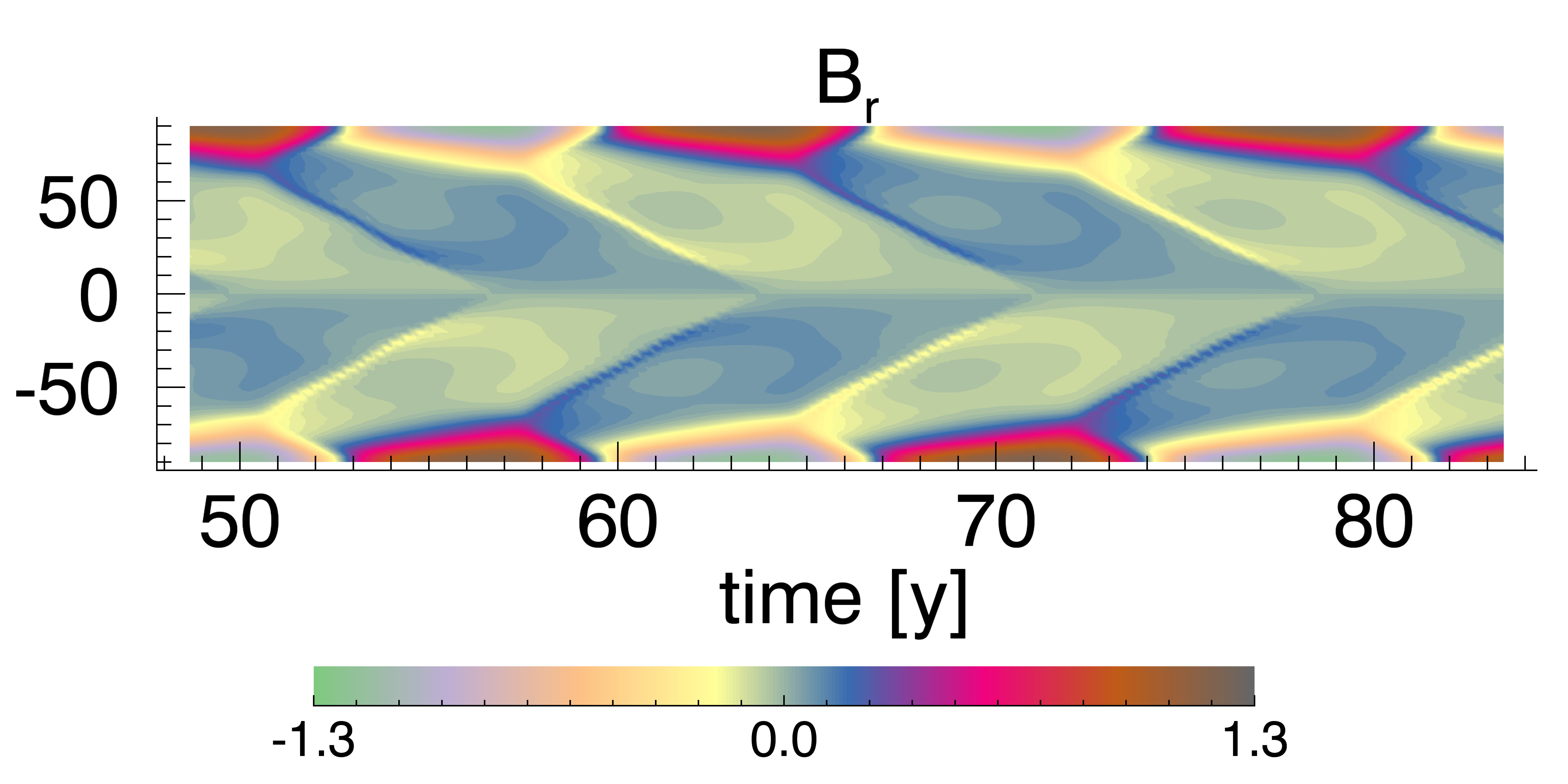}
 \includegraphics[width=2.3in]{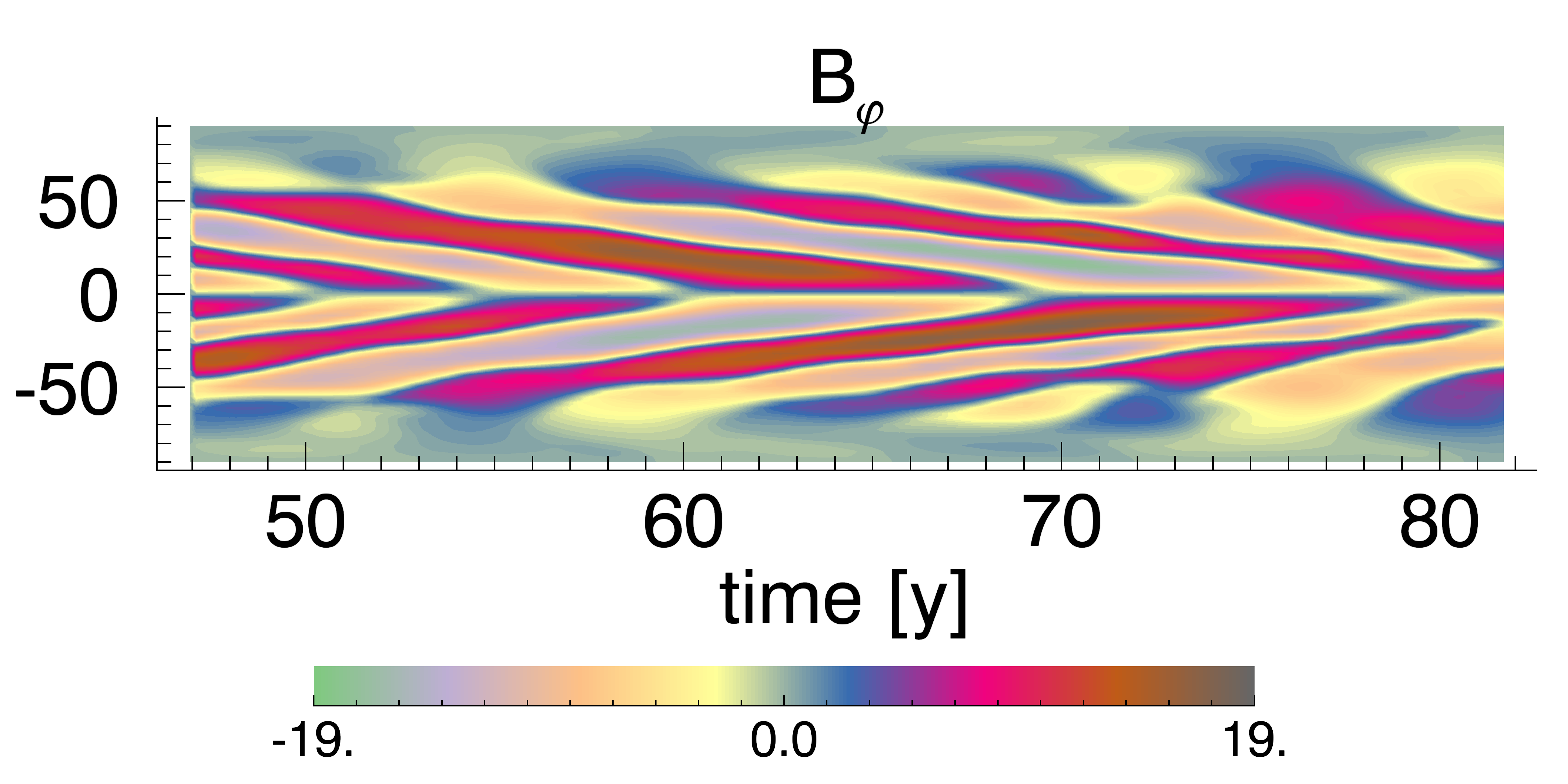}
 \includegraphics[width=2.3in]{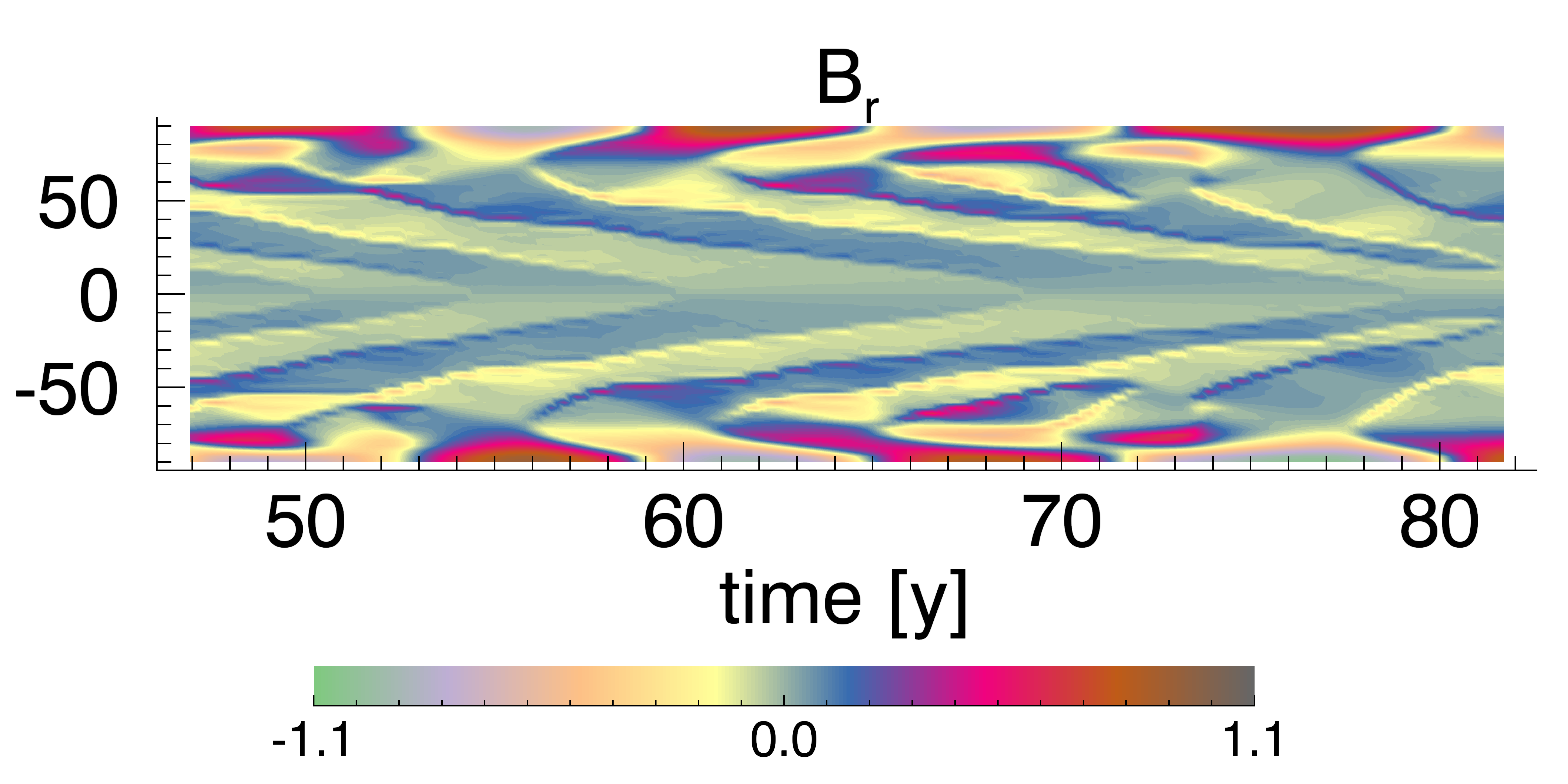}
 \includegraphics[width=2.3in]{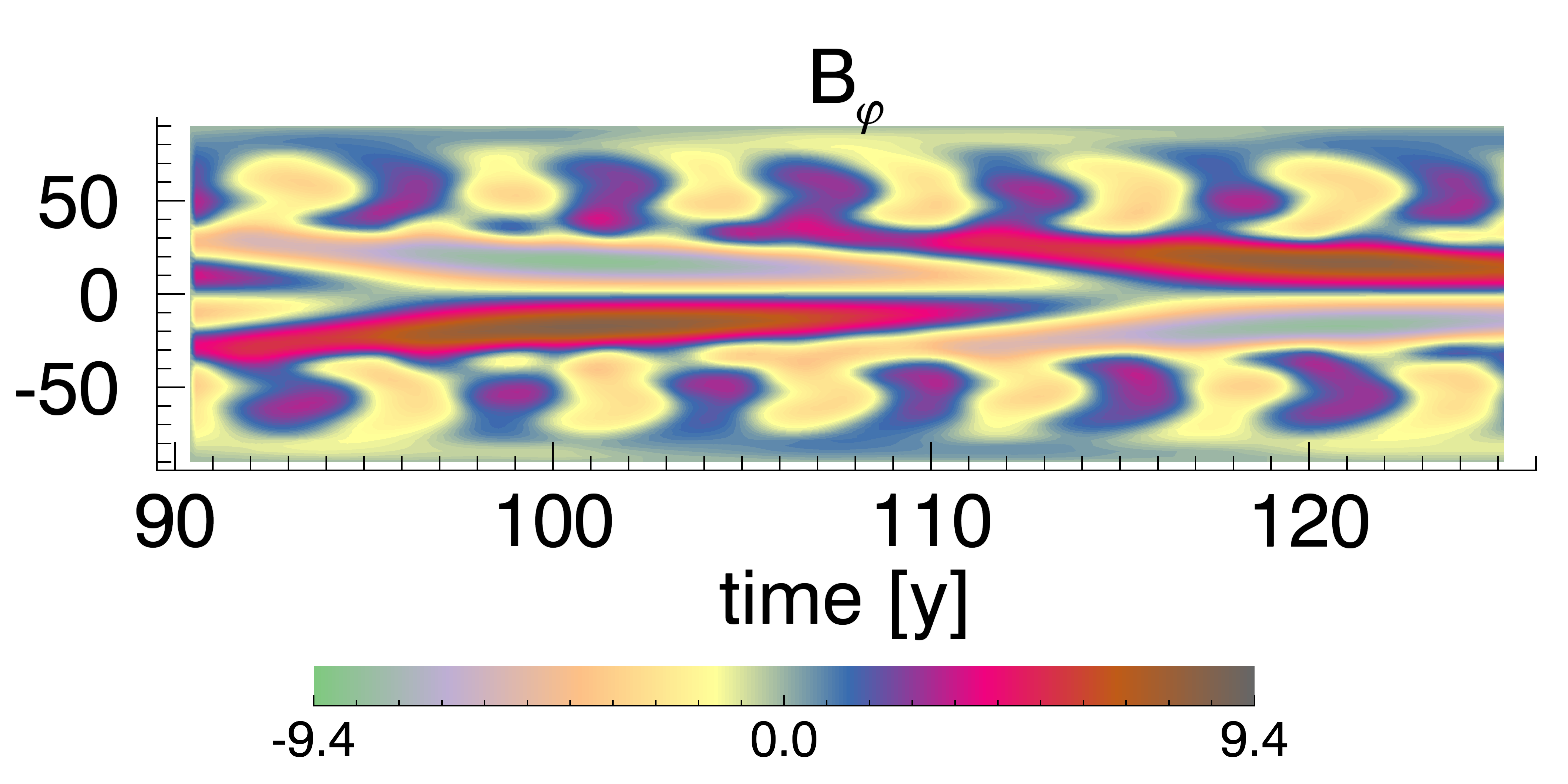}
   \includegraphics[width=2.3in]{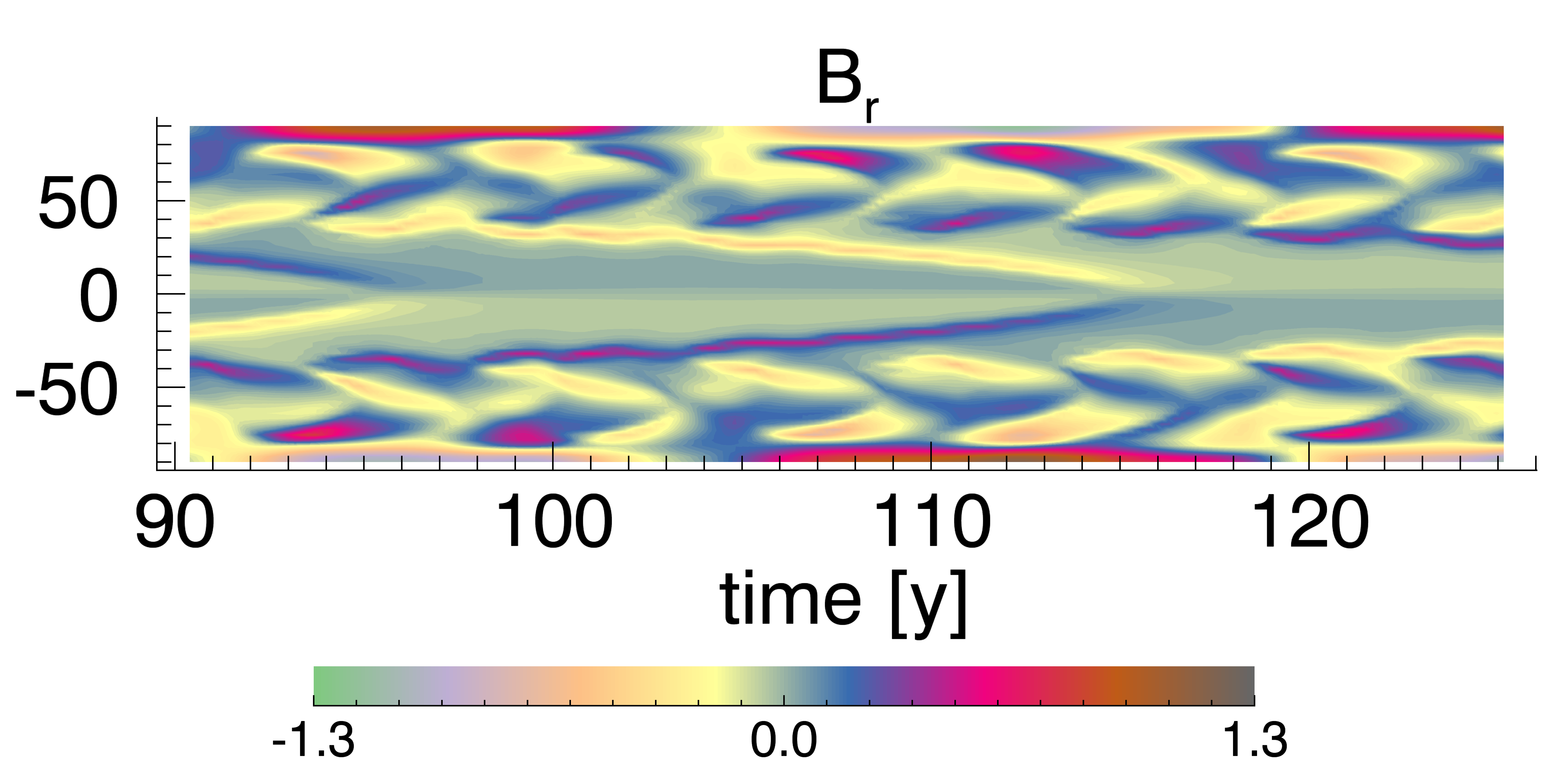}
% \vspace*{-3.0 cm}
 \caption{Time-latitude diagrams for halo diffusivities $\eta_h=10\eta_c$ (upper), $\eta_h=\eta_c$ (middle) and $\eta_h=0.1\eta_c$ (lower). The halo rotates as the pole.}
   \label{fig3}
\end{center}
\end{figure}
We show butterfly diagrams for the toroidal and radial field beneath the solar surface  for a corona rotating
with the angular velocity of the solar surface in Fig.\,\ref{fig1}. 
These models have no radial shear at the solar surface but latitudinal shear in the corona. For a rigidly rotating corona we present two cases, corotation with the core and corotation with the pole in  Fig.\,\ref{fig2} and Fig.\,\ref{fig3}, respectively. 
All the nine models show amazingly different behavior what underlines the role of the boundary condition for the dynamo. 
The models with no radial shear at the stellar surface show no latitudinal migration for the higher diffusivity cases
in the atmosphere, but also an equatorward migration for the low diffusivity. 
But this model has  a mixed mode solution of an axisymmetric oscillation and a strong non-axisymmetric polar field. 
Therefore the spots at the pole in the radial butterfly diagram 
 show the double period caused by the rotation of the mode with azimuthal wave number $m=1$.   
Best models appear for the high diffusivity case of the rigidly rotating atmospherical layer.
Rigid halo rotation with the pole (3) shows a nice butterfly diagram for the toroidal field 
but the equatorward migration of the radial field is better in agreement with the observations of \cite{Hathaway10}
in case (2) of rigid rotation with the core. 

Dynamos with usually taken standard vacuum boundary conditions have no magnetic Poynting flux through the solar surface, whereas it should be different from zero for the magnetic heating of the corona. It is still open how far the large scale field contributes to
the Poynting flux from the photosphere into the chromosphere, which can not be seen in models with a potential field boundary.
Because the large scale dynamo is a global phenomenon it depends crucially on its boundary condition as it is demonstrated with our simple models. How far the rotation in the corona depends on the dynamo can only be found by a dynamical approach (\cite[Warnecke et al.,  2016]{Warnecke16}). Common models of stellar dynamos with their corona are also necessary to explain stellar activity in dependence on rotation.

\section{A dynamo with field dependent memory effect}
In \cite{Fournier18} we found for a Babcock-Leighton type dynamo equatorward migration in the diffusive regime independent of the meridional flow. 
Magnetic flux tubes have a finite rise time scaling with rotation period of the star and the ratio between buoyant force and Coriolis force modified by the magnetic tension. For the sun we have $\tau_{delay} = \tau_0/{\rm sin(\theta) |B_\varphi/B_{eq}|^q}$. It is a non-local $\alpha$-effect in space and time, where the non locality depends on the field strength in a non-linear way for q between -0.91 and -2.0. The toroidal flux can accumulate at the surface during the cycle. It is the  subcritical regime, which leads to the equatorial migration at low latitudes. The system automatically saturates for strong fields, where the accumulation of flux is impossible because of the fast rise time. No additional quenching of the source term is needed. 
\begin{figure}[!htb]
% \vspace*{-2.0 cm}
\begin{center}
%\hspace{-0.5 cm}
 \includegraphics[width=4.7in]{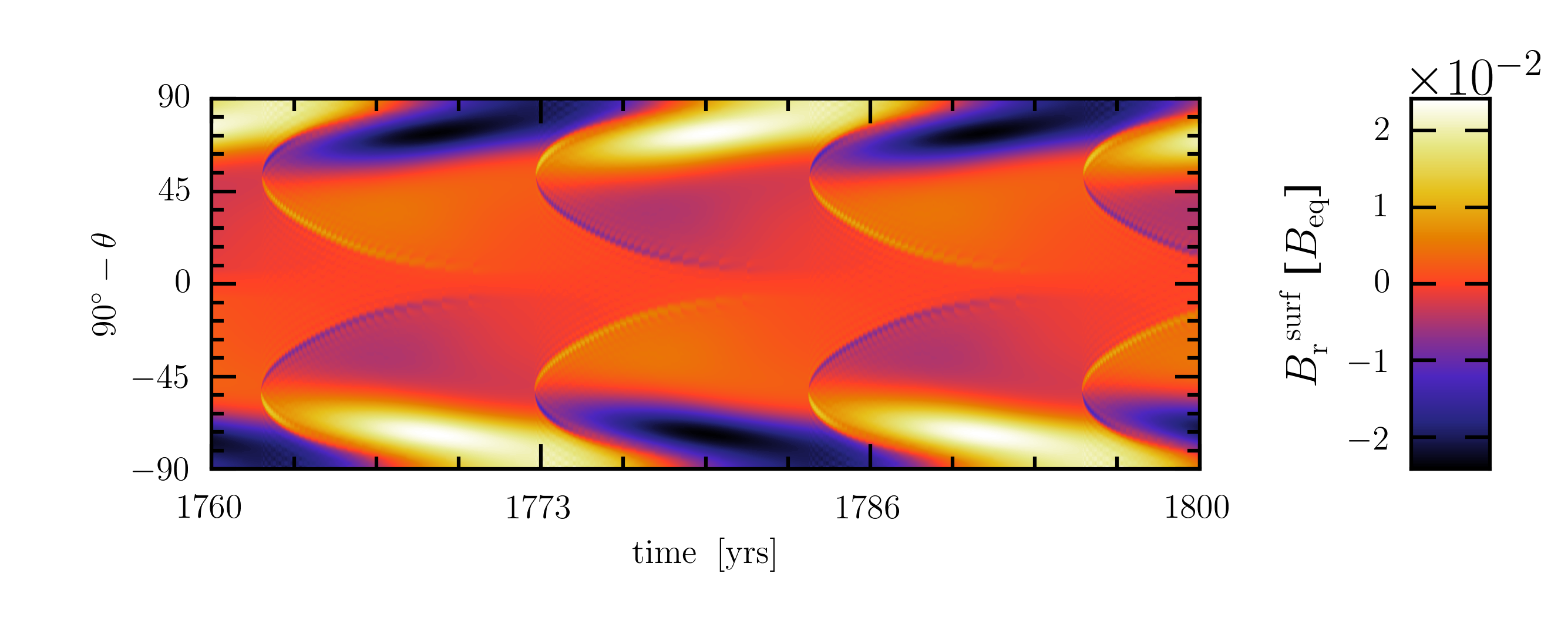} 
 
% \vspace*{-3.0 cm}
 \caption{Butterfly diagram for the radial field of a subcritical Babcock-Leighton dynamo.}
   \label{fig4}
\end{center}
\end{figure}

\end{document}